\documentclass[aps,prb,twocolumn,
	groupedaddress,superscriptaddress,
	amsfonts,amssymb,amsmath,floatfix,
	citeautoscript]{revtex4-1} \usepackage{graphicx} 
\usepackage[centering,hmargin=20mm,tmargin=28mm,bmargin=28mm]{geometry}
\usepackage{amsmath}
\usepackage{hyperref} 
\usepackage{multirow}
\usepackage{newtxtext}                                        
\usepackage{booktabs}
\usepackage{xcolor} 
\usepackage{hyperref}
\usepackage[normalem]{ulem}
\usepackage[textwidth=1.5cm,textsize=footnotesize]{todonotes}
\hypersetup{colorlinks, 
	linkcolor={blue!75!black!80!yellow},
	citecolor={blue!75!black!80!yellow}, 
	urlcolor={blue!75!black!80!yellow}
	}
\usepackage[cmintegrals]{newtxmath}
\newcommand{\BostonCollege}{Department of Physics, Boston College, Chestnut Hill, MA, USA}
\newcommand{\HarvardSEAS}{John A. Paulson School of Engineering and Applied Sciences, Harvard University, Cambridge, MA, USA}
\newcommand{\UCLA}{Department of Physics and Astronomy and California NanoSystems Institute,
University of California, Los Angeles, CA 90095, USA}

\makeatletter
\renewcommand\@make@capt@title[2]{	\@ifx@empty\float@link{\@firstofone}{\expandafter\href\expandafter{\float@link}}	\sffamily{\textbf{#1}}\@caption@fignum@sep#2
}
\makeatother

\begin{document} 
\author{Jennifer Coulter}\email{jcoulter@g.harvard.edu}\affiliation{\HarvardSEAS}
\author{Gavin B. Osterhoudt}\email{osterhog@bc.edu}\affiliation{\BostonCollege}
\author{Christina A. C. Garcia}\affiliation{\HarvardSEAS}
\author{Yiping Wang}\affiliation{\BostonCollege}
\author{Vincent Plisson}\affiliation{\BostonCollege}
\author{Bing Shen}\affiliation{\UCLA}
\author{Ni Ni}\affiliation{\UCLA}
\author{Kenneth S. Burch}\affiliation{\BostonCollege}
\author{Prineha Narang}\email{prineha@seas.harvard.edu}\affiliation{\HarvardSEAS}

\title{Uncovering Electron-Phonon Scattering and Phonon Dynamics in Type-I Weyl Semimetals}
\date{\today}
\begin{abstract}
Weyl semimetals are 3D phases of matter with topologically protected states that have remarkable macroscopic transport behaviors. As phonon dynamics and electron-phonon scattering play a critical role in the electrical and thermal transport, we pursue a fundamental understanding of the origin of these effects in type-I Weyl semimetals NbAs and TaAs. In the temperature-dependent Raman spectra of NbAs, we reveal a previously unreported Fano lineshape, a signature stemming from the electron-phonon interaction. Additionally, the temperature dependence of the $A_{1}$ phonon linewidths in both NbAs and TaAs strongly deviate from the standard model of anharmonic decay. To capture the mechanisms responsible for the observed Fano asymmetry and the atypical phonon linewidth, we present first principles calculations of the phonon self-energy correction due to the electron-phonon interaction. Finally, we investigate the relationship between Fano lineshape, electron-phonon coupling, and locations of the Weyl points in these materials. Through this study of the phonon dynamics and electron-phonon interaction in these Weyl semimetals, we consider specific microscopic pathways which contribute to the nature of their macroscopic transport. 
\end{abstract}
\maketitle

Weyl semimetals host nondegenerate band crossings that are the 3D analog of Dirac points in graphene and are a condensed matter realization of massless fermions described by the Weyl equation\cite{Weyl1929, WeylDiracRMP}. Recently, the transition-metal monopnictides (TaAs, NbAs, TaP, NbP) were predicted to be Weyl semimetals\cite{Weng2015, Huang2015} and experimentally confirmed as such \emph{via} angle-resolved photoemission spectroscopy and quantum oscillation measurements of the characteristic bulk dispersion and surface Fermi arcs\cite{Xu2015_TaAs, Lv2015_TaAs1, Lv2015_TaAs2, Yang2015_TaAs, Xu2015_NbAs, Xu2015_TaP, Xu2015_NbP, Xu2016_TaP, Souma2016_NbP, WeylDiracRMP, Ghimire2015,Zhang2015R,Shekhar2015}. From this discovery sprang reports of the linear and nonlinear optical response of these materials\cite{Xu2016, Wu2017, Kimura2017, Sun2017, Neubauer2018, Patankar2018,osterhoudt2019} as well as many magnetotransport studies seeking definitive signatures of the chiral anomaly\cite{Nielsen1983, Zyuzin2012, Son2013, Parameswaran2014} in this family\cite{Huang2015b, Zhang2015, Hu2016, Arnold2016a, Du2016, Wang2016, Niemann2017, Zhang2017, Sudesh2017} for further confirmation of the Weyl semimetal phase. While the observation of the chiral anomaly is still a subject of active debate\cite{dosReis2016, Li2017, WeylDiracRMP}, reports of these materials' giant magnetoresistance and ultrahigh mobilities make their transport a topic of fundamental importance\cite{Ghimire2015, Zhang2015R, Shekhar2015, Luo2015, Huang2015b, Hu2016, Arnold2016a, Du2016, Wang2016, Niemann2017, Zhang2017, Sudesh2017}.

Yet, despite their critical contributions to electrical transport, the phonon and electron-phonon properties of these materials remain largely unexplored. Therefore, a true understanding of the macroscopic transport in these Weyl semimetals demands a thorough investigation of these effects. In this \emph{Letter} we present an in-depth study of phonon dynamics in the Weyl semimetals NbAs and TaAs through \emph{ab initio} descriptions and experimental temperature-dependent Raman spectroscopy. Through these measurements, we probe the zone-center phonon dynamics and observe a set of previously unreported Fano lineshapes in the $B_{1}$ modes of NbAs as well as anomalous temperature dependence of the $A_{1}$ mode linewidths in both NbAs and TaAs. To characterize the mechanisms responsible for the observed Fano lineshapes and $A_{1}$ linewidths, we present calculations of the electron-phonon coupling in both materials. In particular, we predict the correction to the phonon self-energy as a result of the electron-phonon interaction to better understand how specific phonon modes couple to electronic states in these materials. Stronger electron-phonon coupling is observed in NbAs than in TaAs, which is experimentally consistent with the absence of a Fano lineshape and smaller phonon linewidth. We further discuss how the availability of electron and phonon states could provide channels for electron-phonon scattering, a possible contributor to the high mobility of these materials.

\begin{figure*}
\includegraphics[scale=0.24]{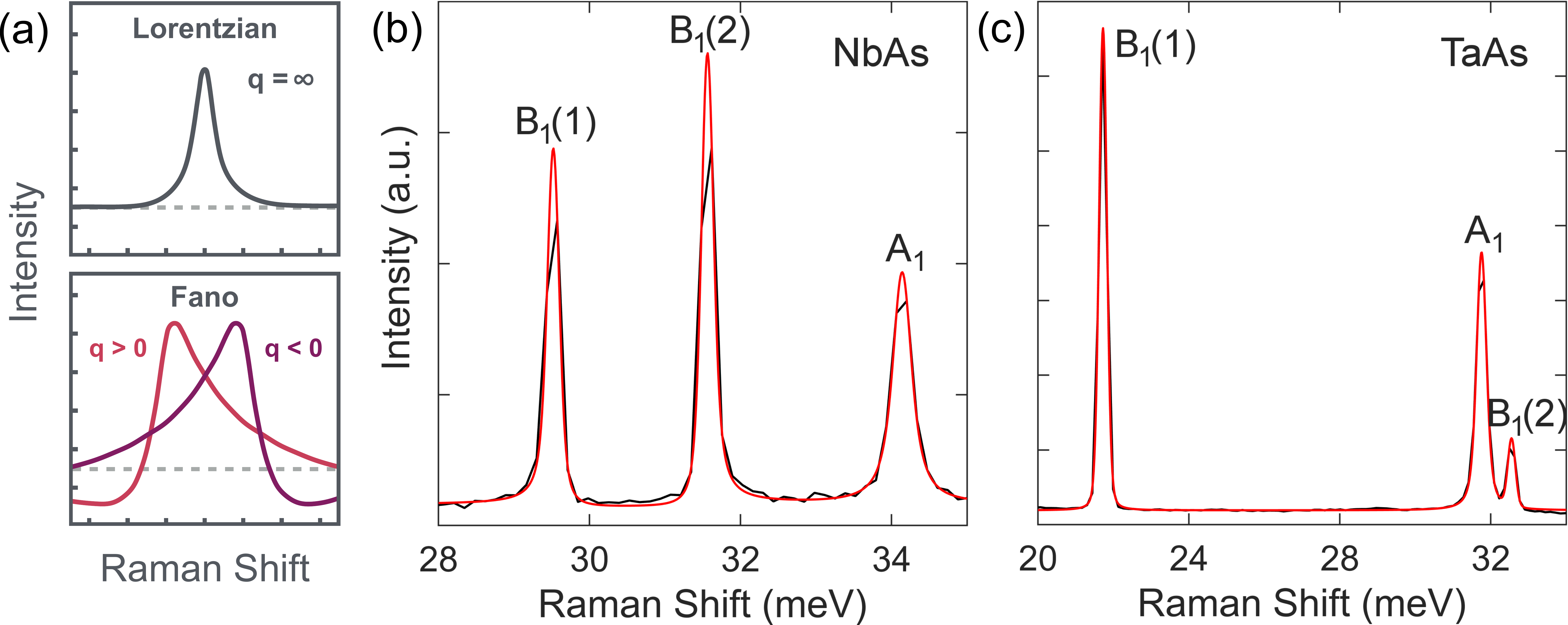}
\caption{Observation of the Fano Lineshape (a) A schematic of the typical Lorentzian and less common Fano lineshapes observed in Raman spectra. (b) Raman spectra of NbAs at 5 K in XX polarization. The $B_{1}$ phonon modes display a subtle and previously unreported Fano lineshape. (c) Raman spectra of TaAs at 10 K in XX polarization. No Fano asymmetry is observed for the $B_{1}$ modes. Red lines are fits to the experimental data.}
\label{fig:1}
\end{figure*}

NbAs and TaAs form in a tetragonal crystal structure with space group $I4_{1}md$ (No. 109) and point group $C_{4v}^{11}$. Their four atom unit cell results in a total of twelve phonon modes, of which three are acoustic and nine optic. The symmetries and selection rules of the optical modes may be determined through the use of group theory, which predicts the existence of one singly-degenerate $A_{1}$ mode, two singly-degenerate $B_{1}$ modes, and three doubly-degenerate $E$ modes\cite{liuPRB15} (corresponding to atomic displacement patterns shown in SI). A previous study\cite{liuJPCM16} confirmed these designations via the polarization dependence of the room temperature Raman spectra of each member of the (Nb,Ta)(P,As) family of compounds. An additional mode was identified in NbAs that is not permitted by crystal symmetry, and thus ascribed to defects.

The Raman spectra at 5 K in NbAs and at 10 K in TaAs, collected from $ab$-surfaces, are shown in Fig.~1(b) and 1(c) respectively (with the full temperature dependence up to 300 K found in the SI). The three phonon modes observable on this surface have been identified by their polarization dependence\cite{liuJPCM16} with symmetries $B_{1}(1)$, $B_{1}(2)$, and $A_{1}$ (in order of increasing energy in NbAs), with the order switched in TaAs to be $B_{1}(1)$, $A_{1}$, and $B_{1}(2)$. All three modes are observable across the entire temperature range investigated and show no evidence of any structural changes, in agreement with previous studies of the (Ta, Nb) (P, As) family of materials\cite{luo2016JPCM}.

Noticeable in the two $B_{1}$ modes of NbAs is the asymmetry present in each of their lineshapes. The asymmetry is explicitly confirmed by comparison of fits to the modes with Gaussian broadened Lorentzian profiles (Voigt profile) and Gaussian broadened Fano lineshapes\cite{schippersJQSRT18}, where the Gaussian convolution accounts for the finite spectral resolution of our spectrometer\cite{tian2D2016}. The Gaussian broadened Fano lineshapes produce a visibly better fit to the experimental data and a comparison of the $\chi^{2}$ metric reveals a quantitative superiority as well (see SI for further details). The asymmetry in each mode is present across the entire temperature range, and notably, the two modes have excess spectral weight on opposite sides, with the lower energy $B_{1}(1)$ mode having more spectral weight on its low energy side and the higher energy $B_{1}(2)$ mode having more spectral weight on the high energy side, seen in Fig.~\ref{fig:1}(b).

\begin{figure*}
\includegraphics[scale=0.28]{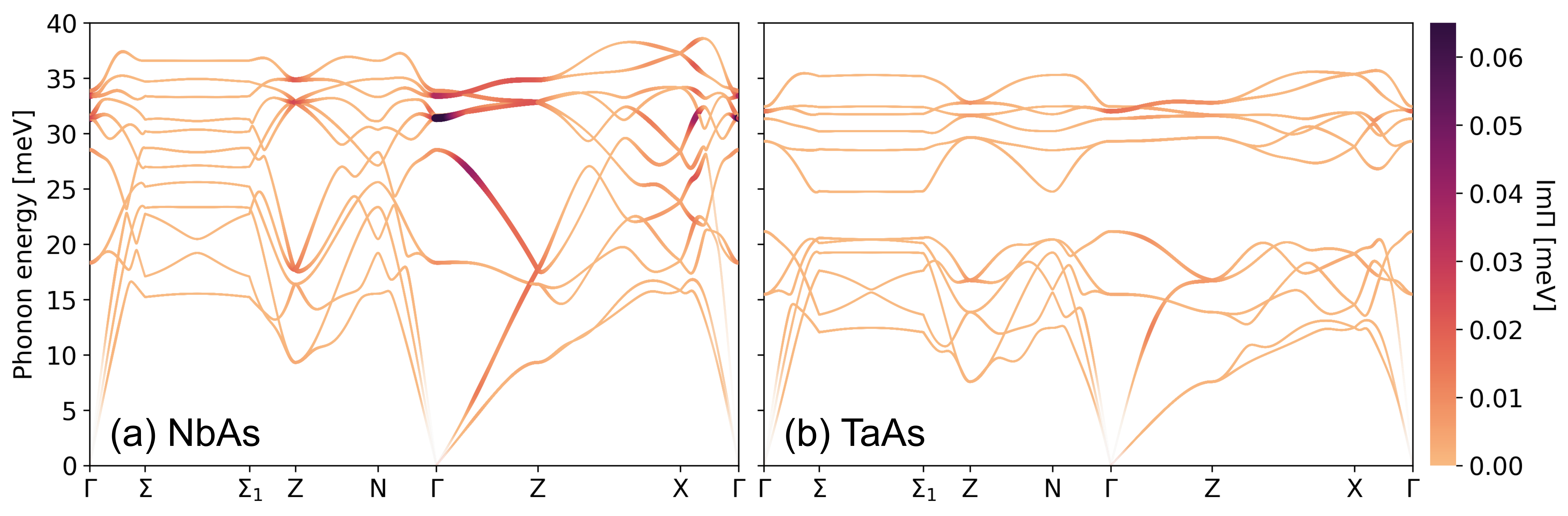}
\caption{Imaginary Component of the Phonon Self-Energy in NbAs and TaAs: The correction to the imaginary part of the phonon self-energy due to the electron-phonon interaction at room temperature projected onto the phonon dispersions of (a) NbAs and (b) TaAs. For both materials, there is a notable increase in self-energy at the $\Gamma$ point, Z point, and along the $\Gamma$-X line, offering insight into which phonon modes and wave vectors play a significant role in macroscopic transport. Though the overall calculated self-energy is less for TaAs than NbAs, similar wave vectors show enhanced phonon self-energy in both materials.}
\label{fig:dispersion}
\end{figure*}

\begin{figure}
    \includegraphics[scale=0.95]{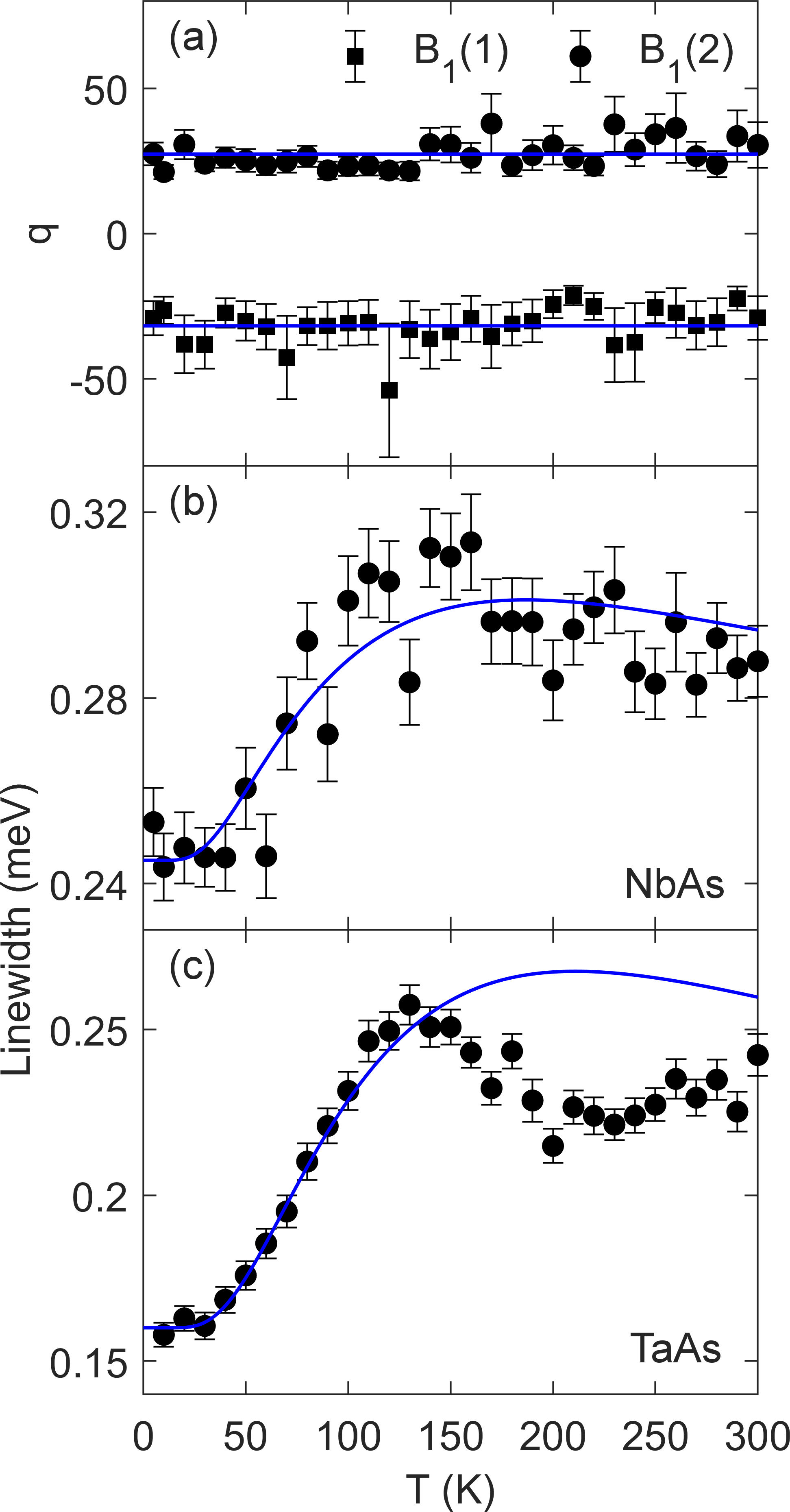}
    \caption{(a) Temperature dependence of the Fano asymmetry $q$ for the two $B_{1}$ modes in NbAs. The blue line represents the average $q$ value across all temperatures. (b) Temperature dependence of the $A_{1}$ phonon mode linewidth in NbAs and (c) TaAs. Fits to the data are based on a model of phonon decay into electron-hole pairs. While this model captures the dependence over the whole temperature range in NbAs, it only describes the low temperature behavior of TaAs.}
    \label{fig:3}
\end{figure}

A Fano lineshape is commonly interpreted as a result of the quantum mechanical interference between a discrete state and a continuum of states\cite{fanoPR61,xuNComm17}. As NbAs is nonmagnetic, the only continuum at these energies for such an interference is electronic. A previous investigation of the infrared reflectance of TaAs has observed a Fano lineshape in the $A_{1}$ phonon mode which was interpreted as arising from coupling to the low energy Weyl fermion-like excitations\cite{xuNComm17}. The asymmetry in the previously reported TaAs measurements was strongly temperature dependent, while in contrast, the asymmetry observed here in NbAs appears to be nearly temperature independent. In Fig.~\ref{fig:3}(a) we show the temperature dependence of the Fano asymmetry parameter $q$ obtained from fitting our spectra. The $q$ for each mode appears to be independent of temperature, and interestingly, the two modes appear to have nearly the same $q$ but of opposite sign with the lower energy $B_{1}(1)$ mode having $q_{\text{avg}} = -31.82 \pm 6.4$ and the higher energy $B_{1}(2)$ mode having $q_{\text{avg}} = 27.36 \pm 4.50 $. In comparison, neither of the $B_{1}$ modes in TaAs display any Fano asymmetry, nor does the $A_{1}$ mode\cite{liuPRB15} in which it was previously observed in IR measurements\cite{xuNComm17}.

While the $A_{1}$ modes in NbAs and TaAs appear to lack any asymmetry in their lineshapes, their linewidths display atypical temperature dependent behavior. As shown in Fig.~\ref{fig:3}(b) and (c), we observe that the $A_{1}$ linewidths undergo a sharp increase at temperatures below $\approx150$ K, followed by a decreasing trend towards room temperature in the case of NbAs, while in TaAs this trend appears to reverse and the width starts to increase once more around $\approx250$ K. The standard anharmonic model of optical phonon decay into two or more acoustic phonons predicts a monotonically increasing temperature dependence\cite{klemens1966} and is unable to describe this observed behavior (though the dependence of all other measured phonons does match this prediction - see SI for further details). We also note that the linewidths of the $A_{1}$ modes are much larger than the other phonons in these materials.

Motivated by these experimental observations, we calculate the electron-phonon coupling in both NbAs and TaAs. To understand the impact of the electron-phonon interaction on a specific phonon mode, it is useful to evaluate the imaginary part of the phonon self-energy correction due to the electron-phonon interaction in the Migdal approximation\cite{giustino2017},
\begin{multline} 
\mathrm{Im}\Pi_{\textbf{q}\alpha} = \frac{\hbar}{2\tau_{\textbf{q}\alpha}} = 2\pi \sum_{mn}  \int_{\mathrm{BZ}} \frac{\Omega d\textbf{k} }{(2\pi)^3} \left| g^{\textbf{q}\alpha}_{\textbf{k} m, (\textbf{k} + \textbf{q})n} \right|^2\\
\times (f_{\textbf{k}n} - f_{(\textbf{k}+\textbf{q})m})  \delta(\hbar \omega_{\textbf{q}\alpha} - \varepsilon_{(\textbf{k} + \textbf{q})m} + \varepsilon_{\textbf{k}n}),
\end{multline}
which is related to the lifetime associated with the scattering of phonons in the electron-phonon interaction and is directly proportional to the corresponding phonon scattering rate. 

To compute this self-energy, we use first-principles density functional theory calculations of the electron-phonon matrix elements from JDFTx\cite{JDFTx}. Using a basis of maximally-localized Wannier functions\cite{giustino2007}, we then interpolate all energies and matrix elements to a much finer $k$-mesh to converge the necessary Brillouin zone integral. Methods for this calculation are presented in our previous work\cite{nanoLett:CC, Coulter:2018, ACSNanoBrown2016,PhysRevB.97.195435}. The independent electron and phonon states were calculated using fully-relativistic ultrasoft pseudopotentials\cite{dal_corso_pseudopotentials_2014,rappe_optimized_1990} parameterized for the PBEsol exchange-correlation functional\cite{PBEsol}.

The calculated imaginary part of the self-energy is plotted along the phonon dispersions in Fig.~\ref{fig:dispersion} to visually portray how the probability of participating in e-ph scattering varies among specific phonon states and between the different materials. From this calculation, we notice an increase in the self-energy of the optical modes near the zone-center, and see that the $B_{1}$ modes both experience some coupling to the electronic system. While the overall imaginary component of the self-energy is moderate (as typical metals like Pb, Cu, Pd, or Al feature energies of approximately 0.01-0.2 meV), the calculated value is reasonable given the relatively weak Fano asymmetry observed in experiment. Nonetheless, given the reduced electronic density of states compared to conventional metals, this could be considered strong electron-phonon coupling relative to the number of available electronic states. We note that similar coupling strength is predicted in all of the optical modes around 30 meV, despite the asymmetry only arising in the $B_{1}$ modes experimentally, and that phonons with wave vectors along the $\Gamma$-X and $\Gamma$-Z lines also have a significant probability of participating in e-ph scattering.

Additionally, we compare these Im$\Pi$ calculations for NbAs and TaAs as in Fig.~\ref{fig:dispersion}. Visually, it is apparent that the electron-phonon intereaction plays a larger role NbAs than in than in TaAs. To quantify this, we also present the self-energy specifically at the $\Gamma$ point for each material and their ratio in SI Table 1. From the ratio of their self-energies, we can see that in general, the effect of the electron-phonon interaction in NbAs is stronger in all phonon modes, consistent with the relative experimental linewidths and the observation of Fano lineshape in NbAs but not TaAs. 

In light of the calculated increase in phonon self-energy at certain wave vectors, we also note the specific phase-space locations of the Weyl points. In the lattice coordinates of the standard conventional unit cell for each material, we find that the Weyl points appear at $\vec{k}_{1}$ = [0.003, 0.475, 0.000] with $E_{1}$ = (-30.4 $\pm$ 0.7) meV and $\vec{k}_{2}$ = [0.007, 0.277, 0.559] with $E_{2}$ = (-20.3 $\pm$ 0.4) meV for NbAs, and $\vec{k}_{1}$ = [0.008, 0.505, 0.000] with $E_{1}$ = (-33.3 $\pm$ 0.3) meV and $\vec{k}_{2}$ = [0.020, 0.280, 0.578] with $E_{2}$ = (-37.2 $\pm$ 0.5) meV for TaAs. Here, 1 and 2 indicate labels for the two unique Weyl point types (hereafter W1 and W2) observed in these materials, which are transformed under the appropriate time-reversal, crystal mirror and planar symmetries to obtain all 24 crossings in each materials' band structure. While the momentum-space coordinates of these Weyl points are in reasonably good agreement with prior works,\cite{Weng2015,Huang2015,Lee2015} there is some discrepancy in our Weyl point energies, with the most notable difference in the energetic positions of W2, which other work claims to be closer to or even above the Fermi energy. We note that regardless of this, these energies are on the same scale as the optical phonon modes for both materials. 

We first discuss the appearance of the Fano lineshape for the $B_{1}$ modes in NbAs but its absence in the $A_{1}$ and $E$ modes. The appearance in modes of only one symmetry, despite several modes of different symmetries occurring with nearly the same energy at $\Gamma$, suggest that it is symmetry -- and not energy -- considerations that are responsible. This is indicated by the near energy degeneracy of the $B_{1}$ and $E$ modes at $\Gamma$\cite{liuJPCM16} (see Fig.~\ref{fig:dispersion} and SI). A requirement for the appearance of Fano asymmetry is the matching of symmetries between the Raman tensor for the phonon and for electronic scattering\cite{cardonaPRB78,cardonaPRB74}. We thus consider the case of electronic Raman scattering in graphene, where the two-step scattering process dominates over the one-step or ``contact'' interaction due to the energy bandwidth $\gamma_{0}$ being larger than the energy $\Omega$ of the laser (cf. Eq.'s (14) and (15) in reference \cite{KashubaNJP2012}). 

Here, we find ourselves in the opposite limit, with the bandwidth of the Weyl nodes in NbAs being much smaller ($\approx 0.1-0.2$ eV) than the energy of the laser used in our experiments ($2.33$ eV). We thus expect that the contact interaction dominates electron-hole pair creation in the Raman scattering process of NbAs. The symmetries available to electron-hole pairs generated by the contact interaction are determined by the specific forms of the second-order corrections to the linear spectrum\cite{KashubaNJP2012}. To the best of our knowledge, this is yet to receive a fully detailed theoretical analysis. Therefore, while we expect that the contact interaction is responsible for the electronic continuum indicated by the observed Fano lineshapes, the details required by symmetry remain an open theoretical question and the subject of future work.

To further understand the role of electron-phonon scattering, we carefully examined the temperature dependence of the $A_{1}$ linewidth. Indeed, since anharmonic decay involves only bosons, its temperature dependence is quite distinct from that caused by electron-phonon scattering. As shown in Fig.~\ref{fig:3}, we find that the non-monotonic temperature dependence of the $A_{1}$ mode linewidth of both NbAs and TaAs is well described by a model of phonon decay into electron-hole pairs. This is consistent with a previous IR work on TaAs,\cite{xuNComm17} though the previously observed linewidth decreased monotonically with temperature. However, as detailed in the SI, we found that by generalizing their model to account for a finite chemical potential and the tilt of the Weyl nodes (breaking particle-hole symmetry) we were able to capture the features present in our measured temperature dependence. Fits to our data using the generalized model are plotted as blue lines in Fig.~\ref{fig:3}(b) and (c). Overall, this model accounts well for the temperature dependence of both materials, though the much larger linewidth in NbAs suggests stronger electron-phonon scattering, consistent with our \textit{ab-initio} studies (Fig.~\ref{fig:dispersion}). This is further confirmed as the temperature dependence of the mode energy, indicates similar anharmonicity. The deviations of the data from the model in TaAs likely result from contributions of the W1 node, as it is only $\approx 14$ meV\cite{Xu2015_TaAs} below the W2 node, whereas it is considerably lower ($\approx 38$ meV\cite{Xu2015_NbAs}) in NbAs.

\begin{figure}[h!]
\includegraphics[scale=0.23]{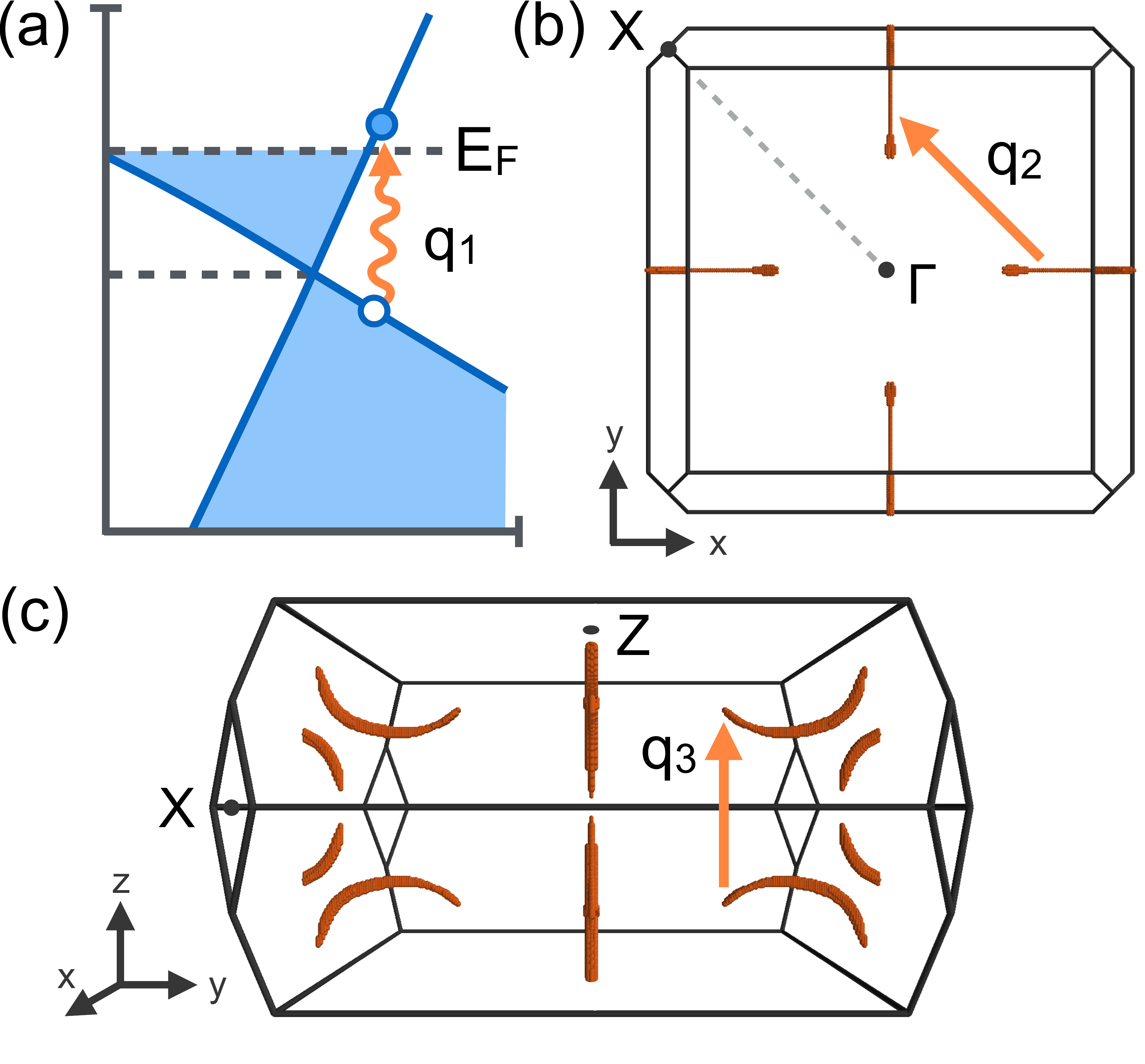}
\caption{(b) and (c) show the Brillouin zone of NbAs, with areas in which any two bands are within 31meV of both each other and $E_F$ are shown in orange. From this, we can see correspondence between areas of the zone which are connected by specific phonon wave vectors and areas of the phonon dispersion which are more likely to participate in e-ph scattering. (a) Shows a possible transition within a single Weyl cone. In these images, q$_1$ is a wave vector near $\Gamma$, and q$_2$ and q$_3$ are respectively along the lines from $\Gamma$-X and $\Gamma$-Z.}
\label{fig:4}
\end{figure}

Having established phonon decay into electron-hole pairs as a primary scattering mechanism in the $A_1$ modes, we aim to understand the specific scattering pathways which contribute to electron-phonon scattering. Specifically, we identify particular wave vectors in the phonon dispersion that experience enhanced phonon self-energy by considering the electronic states available for scattering via phonons. As in Fig.~\ref{fig:4}, we can select electronic states in the Brillouin zone which are within 31 meV of another state, where both states within 31 meV of $E_{F}$, to visualize the electronic states available for electron-hole generation by a phonon in these materials. 

We find that the possible states do correspond well to the wave vectors that display enhanced self-energy along the phonon dispersion. In Fig.~\ref{fig:4}, we show possible transitions. In Fig.~\ref{fig:4}(a), we show transitions due to $\vec{q}_1$, which represents a wave vector on or very near to $\Gamma$, which could scatter between states within the same Weyl cone or across very closely neighboring pairs. In Fig.~\ref{fig:4}(b), we show a possible $\vec{q}_2$, connecting $k$-space locations diagonally across the zone within the same $k_{z}$ plane via a wave vector along the $\Gamma$-X line. In Fig.~\ref{fig:4}(c), we indicate $\vec{q}_3$, representing transitions in the $k_{z}$ direction along the $\Gamma$-Z line nearly equal to the distance from $\Gamma$ to Z. These wave vectors match well to the ones that were seen in the self-energy calculation, and in particular for wave vectors near the $\Gamma$-Z line and $\Gamma$ point, correspond to the $k$-space distances which connect areas surrounding Weyl cones. Taken together, these results suggest the existence of specific inter- and intra-Weyl node scattering pathways available via phonons. 

In conclusion, our combined theoretical and experimental study identifies important features of the phonon and electron-phonon properties of these type-I Weyl semimetals. From first principles calculations of the electron-phonon coupling, we predict the imaginary part of the phonon self-energy due to the electron-phonon interaction to uncover new information about specific channels for scattering in NbAs and TaAs. A detailed experimental study of temperature-dependent Raman spectra of NbAs reveals a previously unreported Fano lineshape in the two $B_{1}$ zone-center phonon modes and a temperature dependence of the $A_{1}$ linewidths in both NbAs and TaAs which deviates strongly from traditional anharmonic phonon models. These complimentary predictions and experimental measurements elucidate the role of the electron-phonon interaction and provide a fundamental understanding of the microscopic scattering processes underlying transport in these type-I Weyl materials.

\section*{Acknowledgements}
This work was supported by the DOE Photonics at Thermodynamic Limits Energy Frontier Research Center under Grant No. DE-SC0019140. Analysis and measurements performed by G.B.O. and work done by K.S.B. was supported by the U.S. Department of Energy (DOE), Office of Science, Office of Basic Energy Sciences under Award No. DE-SC0018675. Raman experiments by Y.W. and V.P. were supported by the National Science Foundation through grant DMR-1709987. This research used resources of the National Energy Research Scientific Computing Center, a DOE Office of Science User Facility supported by the Office of Science of the U.S. Department of Energy under Contract No. DE-AC02-05CH11231, as well as resources at the Research Computing Group at Harvard University. J.C. recognizes the support of the DOE Computational Science Graduate Fellowship (CSGF) under Grant No. DE-FG02-97ER25308. C.A.C.G. is supported by the NSF Graduate Research Fellowship Program. Work at UCLA was supported by the U.S. Department of Energy (DOE), Office of Science, Office of Basic Energy Sciences under Award Number DE-SC0011978.

\noindent J.C. and G.B.O. contributed equally to this work.

\end{document}